\documentstyle[12pt,epsfig,axodraw,a4]{article} 
\newcommand{\vs}{\vspace{-0.25cm}}
\begin{document} 

\begin{center}
{\Large{\bf Spin-asymmetry energy of nuclear matter}}

\bigskip

N. Kaiser\\

\medskip

{\small Physik-Department T39, Technische Universit\"{a}t M\"{u}nchen,
    D-85747 Garching, Germany}

\end{center}

\medskip

\begin{abstract}
We calculate the density-dependent spin-asymmetry energy $S(k_f)$ of
isospin-symmetric nuclear matter in the three-loop approximation of chiral 
perturbation theory. The interaction contributions to $S(k_f)$ originate from 
one-pion exchange, iterated one-pion exchange, and (irreducible) two-pion 
exchange with no, single, and double virtual $\Delta$-isobar excitation. We
find that the truncation to $1\pi$-exchange and iterated $1\pi$-exchange terms
(which leads already to a good nuclear matter equation of state) is 
spin-unstable, since $S(k_{f0})<0$. The inclusion of the chiral 
$\pi N\Delta$-dynamics guarantees the spin-stability of nuclear matter. The 
corresponding spin-asymmetry energy $S(k_f)$ stays positive within a wide 
range of an undetermined short-range parameter $S_5$ (which we also estimate 
from realistic NN-potentials). Our results reemphasize the important role 
played by two-pion exchange with virtual $\Delta$-isobar excitation for the 
nuclear matter many-body problem. Its explicit inclusion is essential in order 
to obtain good bulk and single-particle properties.

\end{abstract}

\bigskip



\vspace{1cm}

In recent years a novel approach to the nuclear matter problem based on
effective field theory (in particular chiral perturbation theory) has
emerged. The key element there is a separation of long- and short-distance
dynamics and an ordering scheme in powers of small momenta. At nuclear matter
saturation density $\rho_0 \simeq 0.16\,$fm$^{-3}$ the Fermi momentum $k_{f0}$ 
and the pion mass $m_\pi$ are comparable scales ($k_{f0}\simeq 2 m_\pi$), and
therefore pions must be included as explicit degrees of freedom in the
description of the nuclear many-body dynamics. The contributions to the energy
per particle $\bar E(k_f)$ of isospin-symmetric (spin-saturated) nuclear 
matter as they originate from chiral pion-nucleon dynamics have been computed 
up to three-loop order in refs.\cite{lutz,nucmat}. Both calculations are able
to reproduce correctly the empirical saturation point of nuclear matter by 
adjusting one single parameter (either a coupling $g_0+g_1 \simeq 3.23$
\cite{lutz} or a cut-off $\Lambda \simeq 0.65\,$GeV \cite{nucmat}) related to 
unresolved short-distance dynamics.\footnote{The cut-off scale $\Lambda$ serves
the purpose to tune the strength of an attractive zero-range NN-contact
interaction.} The novel mechanism for saturation in
these approaches is a repulsive contribution to the energy per particle $\bar
E(k_f)$ generated by Pauli-blocking in second order (iterated) one-pion
exchange. As outlined in section 2.5 of ref.\cite{nucmat} this mechanism
becomes particularly transparent by taking the chiral limit $m_\pi = 0$. In
that case the interaction contributions to  $\bar E(k_f)$ are completely
summarized by an attractive $k_f^3$-term and a repulsive $k_f^4$-term where
the parameter-free prediction for the coefficient of the latter is very close
to the one extracted from a realistic nuclear matter equation of state.  

In a recent work \cite{deltamat} we have extended the chiral approach to 
nuclear matter by including systematically the effects from two-pion exchange 
with single and double virtual $\Delta(1232)$-isobar excitation. The physical
motivation for such an extension is threefold. First, the spin-isospin-3/2 
$\Delta(1232)$-resonance is the most prominent feature of low-energy 
$\pi N$-scattering. Secondly, it is well known that the two-pion exchange 
between nucleons with excitation of virtual $\Delta$-isobars generates the 
needed isoscalar central NN-attraction \cite{gerst} which in phenomenological 
one-boson exchange models is often simulated by a fictitious  scalar 
''$\sigma$''-meson exchange. Thirdly, the delta-nucleon mass splitting of 
$\Delta = 293\,$MeV is of the same size as the Fermi momentum $k_{f0} \simeq 
2m_\pi$ at nuclear matter saturation density and therefore pions and 
$\Delta$-isobars should both be treated as explicit degrees of freedom. A
large variety of nuclear matter properties has been investigated in this
extended framework in ref.\cite{deltamat}. It has been found that the
inclusion of the chiral $\pi N \Delta$-dynamics is able to remove most of the 
shortcomings of previous chiral calculations of nuclear matter
\cite{nucmat,pot,liquidgas,lutzcontra}. For example, the momentum-dependence 
of the (real) single-particle potential $U(p,k_f)$ near the Fermi surface 
$p=k_f$ improves significantly. As a consequence of that the critical 
temperature of the first-order liquid-gas phase transition of
isospin-symmetric nuclear matter gets lowered to the realistic value $T_c 
\simeq 15\,$MeV. The isospin properties of nuclear matter improve also
substantially when including the chiral $\pi N\Delta$-dynamics. Instead of
bending downward above $\rho_0$ as in previous chiral calculations
\cite{nucmat,lutzcontra}, the energy per particle of pure neutron matter 
$\bar E_n(k_n)$ and the (isospin) asymmetry energy $A(k_f)$ grow now 
monotonically with density (see Figs.\,10,11 in ref.\cite{deltamat}). Good 
agreement with results of sophisticated many-body calculations and
(semi)-empirical values has been found in the density regime  $\rho=
2\rho_n<0.2\,$fm$^{-3}$ relevant for conventional nuclear physics. 

Given all that success of chiral perturbation theory for the nuclear matter
many-body problem it is nevertheless still necessary to check the 
spin-stability of nuclear matter in that framework. Such an investigation is 
the subject of the present paper. We remind that in the past the criterion of 
spin-stability of infinite nuclear matter has required modifications of 
several phenomenologically very successful Skyrme forces 
\cite{chang,baeckman,krewald}. For recent work on generalized symmetry
energy coefficients in the context of phenomenological Skyrme forces, see also 
ref.\cite{braghin}.

Let us begin with defining the spin-asymmetry energy $S(k_f)$ of infinite 
nuclear matter. Consider this homogeneous many-nucleon system in a partially 
spin-polarized state such that the Fermi seas of the spin-up and spin-down 
nucleons are filled unequally high. With the help of the spin-projection 
operators $(1\pm \sigma_3)/2$ such a  spin-unsaturated configuration is
realized by the substitution:    
\begin{equation} \theta(k_f-|\vec p\,|) \quad \to \quad {1+\sigma_3 \over 2}\, 
\theta(k_{\uparrow}-|\vec p\,|) +{1-\sigma_3 \over 2}\, \theta(k_\downarrow-
|\vec p\,|)\,, \end{equation}
in the medium insertion.\footnote{Medium insertion is a technical notation for
the difference between the in-medium and vacuum nucleon propagator. For further
details, see section 2 in ref.\cite{nucmat}.} Here, $k_\uparrow$ and  
$k_\downarrow$ denote the different Fermi momenta of the spin-up and spin-down
nucleons. Choosing  $k_{\uparrow,\downarrow} = k_f( 1\pm \eta)^{1/3}$ (with 
$\eta$ a small parameter) the total nucleon density $\rho = (k^3_\uparrow +
k^3_\downarrow)/3\pi^2 = 2k_f^3/3\pi^2$ stays constant. The expansion of the 
energy per particle of spin-polarized (isospin-symmetric) nuclear matter:   
\begin{equation} \bar E_{\rm pol}(k_\uparrow,k_\downarrow)= \bar E(k_f) + 
\eta^2\, S(k_f)+ {\cal O}(\eta^4) \,, \qquad  k_{\uparrow,\downarrow} = k_f(
1\pm \eta)^{1/3} \,,  \end{equation}
around the spin-saturation line ($k_\uparrow = k_\downarrow$ or $\eta = 0$) 
defines the spin-asymmetry energy $S(k_f)$. The obvious criterion for the
spin-stability of nuclear matter is then the positivity of the spin-asymmetry 
energy, $S(k_f)>0$. The energy per particle at fixed nucleon density $\rho$ 
must take on its absolute minimum value in the spin-saturated configuration.   

The first contribution to the spin-asymmetry energy $S(k_f)$ comes from the
kinetic energy $\sqrt{M^2+p^2}-M$ of a non-interacting relativistic Fermi gas:
\begin{equation} S(k_f)={k_f^2\over 6 M}-{k_f^4\over 12 M^3}\,,\end{equation}
with $M=939\,$MeV the (average) nucleon mass. The next term in this series, 
$k_f^6/16M^5$, is negligibly small at the densities of interest. 

\begin{figure}
\begin{center}
\includegraphics[scale=1.,clip]{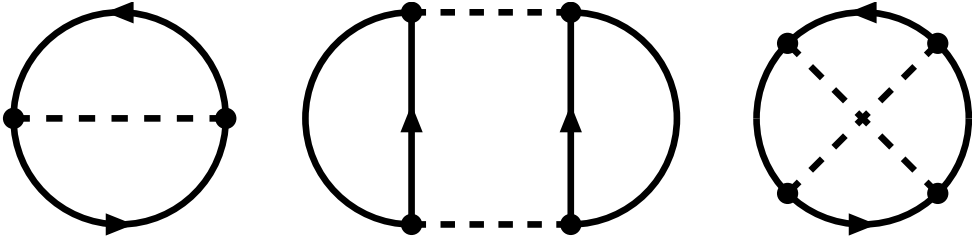}
\end{center}\vspace{-0.4cm}
\caption{The two-loop one-pion exchange Fock diagram and the three-loop
iterated one-pion exchange Hartree and Fock diagrams. The combinatoric factors
of these diagrams are $1/2$, $1/4$ and $1/4$, in the order shown. Their isospin
factors for isospin-symmetric nuclear matter are $6$, $12$ and $-6$,
respectively.}  
\end{figure}

Next, we come to interaction contributions to $S(k_f)$. The closed in-medium 
diagrams related to one-pion exchange (Fock diagram) and iterated one-pion
exchange (Hartree and Fock diagrams) are shown in Fig.\,1. Differences in 
comparison to the calculation of the energy per particle $\bar E(k_f)$ in
ref.\cite{nucmat} occur only with respect to the factors emerging from the
spin-traces over closed nucleon lines and the radii $k_{\uparrow,\downarrow}=
k_f(1\pm \eta)^{1/3}$ of the Fermi spheres to be integrated over. After some
analytical calculation we find the following contribution to the
spin-asymmetry energy $S(k_f)$ from the $1\pi$-exchange Fock diagram in
Fig.\,1 (including its relativistic $1/M^2$-correction):  
\begin{eqnarray} S(k_f) &=& {g_A^2m_\pi^3 \over(4\pi f_\pi)^2} 
\bigg\{\bigg( {u\over 3}+{1\over 8u}\bigg) \ln(1+4u^2)-{u\over 2}-{u^3\over 3} 
\nonumber \\ && +{m_\pi^2\over M^2} \bigg[{5u^3\over 6}-{u^2\over 2} 
\arctan 2u +{u \over 24} (1-6u^2)\ln(1+4u^2)\bigg] \bigg\}\,. \end{eqnarray} 
Here, we have introduced the useful abbreviation $u=k_f/m_\pi$ where $m_\pi 
=135\,$MeV stands for the (neutral) pion mass. As usual $f_\pi = 92.4\,$MeV 
denotes the weak pion decay constant and we choose the value $g_A=1.3$ of the
nucleon axial vector coupling constant in order to have a pion-nucleon 
coupling constant of $g_{\pi N} = g_A M/f_\pi = 13.2$ \cite{pavan}. In the 
second and third diagram in Fig.\,1 the $1\pi$-exchange interaction is 
iterated (once) with itself. These second order diagrams carry the large scale
enhancement factor $M$ (the nucleon mass). It stems from an energy denominator
that is equal to a difference of small nucleon kinetic energies. With a medium
insertion at each of two equally oriented nucleon propagators we obtain from
the three-loop Hartree diagram in Fig.\,1 the following contribution to the
spin-asymmetry energy:  
\begin{equation} S(k_f)= {\pi g_A^4 M m_\pi^4 \over 6(4\pi f_\pi)^4}\bigg\{
\bigg( 15u +{7\over 2u}\bigg) \ln(1+4u^2)-14u -16u^2 \arctan 2u \bigg\} \,. 
\end{equation}
The right Fock diagram of iterated $1\pi$-exchange (see Fig.\,1) with two
medium insertions on non-neighboring nucleon propagators gives rise on the 
other hand to a  contribution to the spin-asymmetry energy of the form:  
\begin{eqnarray} S(k_f) &=& {\pi g_A^4 M m_\pi^4 \over 3(4\pi f_\pi)^4}\bigg\{
{32u^3\over 15} +{7u \over 5} +\bigg( {11 \over 10 u}+{2u \over 3} \bigg)
\ln(1+u^2) -\bigg( {1\over u}+2u\bigg) \ln(1+4u^2)  \nonumber \\  && +\bigg( 
{32 u^4\over 15}-3 \bigg) \arctan u  + (3+4u^2) \bigg[ \arctan
2u + \int_0^u \!dx\, { \arctan x-\arctan 2x \over u(1+2x^2)} \bigg] \bigg\}\,. 
\end{eqnarray}
This expression does not include the contribution of a linear divergence
$\int^\infty_0 dl\,1$ of the momentum-space loop-integral. In dimensional 
regularization such a linear divergence is set to zero, whereas in cut-off 
regularization it is equal to a momentum space cut-off $\Lambda$. The 
additional term specific for cut-off regularization will be given in eq.(13). 
An in-medium diagram with three medium insertions represents Pauli-blocking 
effects in intermediate NN-states induced by the filled Fermi sea of
nucleons. The unequal filling of the spin-up and spin-down Fermi seas shows its
consequences in the spin-asymmetry energy. After some extensive
algebraic manipulations we end up with the following double-integral
representation of the contribution to the spin-asymmetry energy $S(k_f)$ from 
the Hartree diagram in Fig.\,1 with three medium insertions:  
\begin{eqnarray} S(k_f)&=&{g_A^4 M m_\pi^4 \over (4\pi f_\pi)^4 u^3} \int_0^u
\! dx \,x^2 \int_{-1}^1 \! dy \bigg\{ \bigg[{2uxy(3u^2-5x^2y^2) \over (u^2-
x^2y^2)} -(u^2+5x^2y^2)H\bigg] \nonumber \\ && \times \bigg[{2s^2+s^4
\over 1+s^2}-2\ln(1+s^2) \bigg]  +{4u^2 H \,s^5(4s'-3s) \over 3(1+s^2)^2}+ 
\Big[2uxy+(u^2-x^2y^2) H\Big] \nonumber \\ && \times \Big[(5+s^2)(3s^2-8s s' 
+8 s'^2)+8s(1+s^2)(s''-5s'+3s)\Big] {s^4\over 3(1+s^2)^3} 
\bigg\} \,, \end{eqnarray}
where we have introduced several auxiliary functions:
\begin{equation} H = \ln{u+x y\over u- xy}\,, \qquad s= xy
+\sqrt{u^2-x^2+x^2y^2}\,, \qquad s' = u \,{\partial s \over \partial u} \,, 
\qquad s'' = u^2 \, {\partial^2 s \over \partial u^2} \,. \end{equation}
Note that eq.(7) stems from a nine-dimensional principal-value integral over 
the product of three Fermi spheres of varying radii $k_{\uparrow,\downarrow} =
k_f(1\pm \eta)^{1/3}$ which has been differentiated twice with respect to
$\eta$ at $\eta=0$. Of similar structure is the contribution to $S(k_f)$ from 
the  iterated $1\pi$-exchange Fock diagram with three medium insertions. 
Because of the two different pion propagators in the Fock diagram one
ends up (partially) with a triple-integral representation for its contribution
to the spin-asymmetry energy:       
\begin{eqnarray}  S(k_f) &=&{g_A^4Mm_\pi^4\over 24(4\pi f_\pi)^4 u^3}\int_0^u\!
dx\Bigg\{G(3G_{20}-2G_{11}+3G_{02}-8G_{01}-3G) \nonumber \\ && +3G^2_{10} 
-2G_{01}G_{10}+ G_{01}^2 +4x^2 \int_{-1}^1\!dy \int_{-1}^1 \!dz {yz \,\theta
(y^2+z^2-1) \over |yz| \sqrt{y^2+z^2-1}} \nonumber \\ && \times \bigg[{2s^3t^3 
(8s' t-3st -4s't') \over (1+s^2)(1+t^2)}+ {s^2[t^2-\ln(1+t^2)]\over (1+s^2)^2} 
\nonumber \\ && \times \Big[(3+s^2)(8ss'-3s^2-8s'^2)+4s (1+s^2)(6s'-3s-2s'')
\Big] \bigg]\Bigg\} \,.  \end{eqnarray}
Here, we have split into ''factorizable'' and ''non-factorizable'' parts. 
These two pieces are distinguished by whether the (remaining) nucleon 
propagator in the three-loop Fock diagram can be canceled or not by terms from
the product of $\pi N$-interaction vertices. The factorizable terms can be 
expressed through the auxiliary function:   
\begin{equation} G = u(1+u^2+x^2) -{1\over 4x}\big[1+(u+x)^2\big] \big[1+
(u-x)^2\big] \ln{1+(u+x)^2\over 1+(u-x)^2 } \,,   \end{equation}
and its partial derivatives for which we have introduced a (short-hand) 
double-index notation: 
\begin{equation}  G_{ij} = x^i u^j  {\partial^{i+j}G \over \partial x^i 
\partial u^j} \,, \quad 1\leq i+j \leq 2\,. \end{equation}
For the presentation of the nonfactorizable terms one needs also copies of the
quantities $s$ and $s'$ defined in eq.(8) which depend (instead of $y$) on
another directional cosine $z$:  
\begin{equation} t= xz +\sqrt{u^2-x^2+x^2z^2}\,, \qquad t' = u\, {\partial t 
\over \partial u} \,. \end{equation}
For the numerical evaluation of the $dydz$-double integral in eq.(9) it is
advantageous to first antisymmetrize the integrand both in $y$ and $z$ and
then to substitute $z =\sqrt{y^2 \zeta^2 +1-y^2}$. This way the integration
region becomes equal to the unit-square $0< y,\zeta <1$. In the chiral limit
$m_\pi = 0$ the fourth order contributions in eqs.(5-9) sum up to a negative 
$k_f^4$-term of the form: $S(k_f)|_{m_\pi =0} = (g_Ak_f/4\pi f_\pi)^4 (M/135)
(16\pi^2+273-936 \ln2)$. The formulas for the contributions to the 
spin-asymmetry energy $S(k_f)$ written down here in eqs.(3-9) have a strong 
similarity with the analogous contributions to the isospin-asymmetry energy 
$A(k_f)$ collected in eqs.(20-26) of ref.\cite{nucmat}. This is not surprising 
in view of the spin-isospin structure  $\vec \sigma \cdot \vec\nabla \,\tau_a$
of the pion-nucleon coupling. Finally, we give the expression for the linear
divergence specific to cut-off regularization:   
\begin{equation} S(k_f) = - {2 g_A^4 M \Lambda \over (4\pi f_\pi)^4}\, k_f^3 
\,, \end{equation} 
to which only the iterated $1\pi$-exchange Fock diagram (with two medium 
insertions) has contributed. In the case of the Hartree diagram the linear
divergence drops out after taking the second derivative with respect to 
$\eta$.

\begin{figure}
\begin{center}
\includegraphics[scale=0.7,clip]{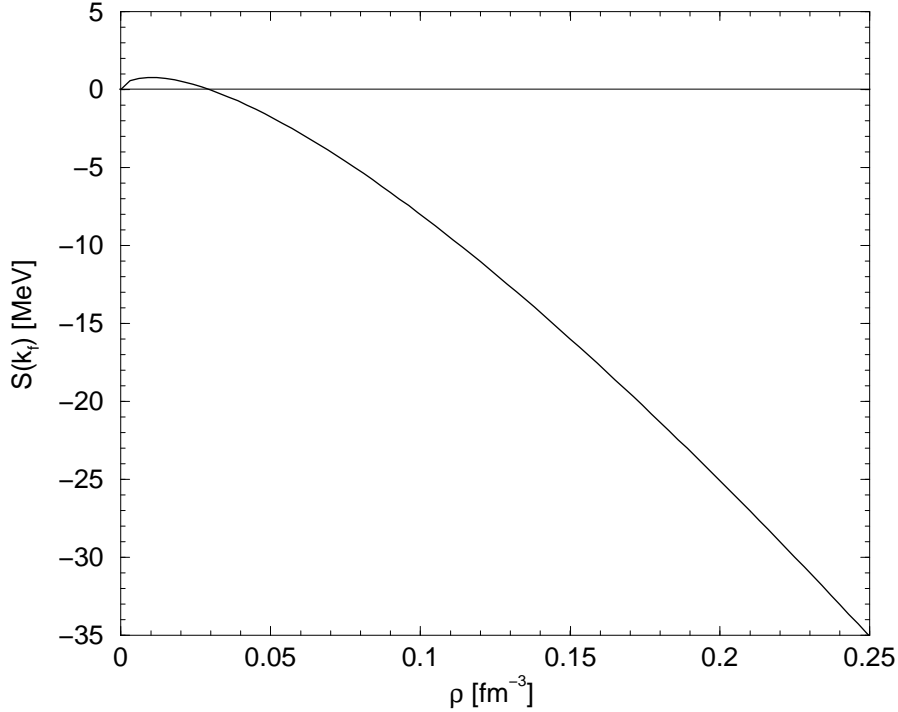}
\end{center}\vspace{-0.8cm}
\caption{The spin-asymmetry energy $S(k_f)$ of nuclear matter versus the 
nucleon density $\rho= 2k_f^3/3\pi^2$. The full line shows the result of a
calculation up to fourth order in small momenta including $1\pi$-exchange
and iterated $1\pi$-exchange. The cut-off scale $\Lambda = 0.61\,$GeV has
been adjusted to the saturation point: $\rho_0 = 0.173\,$fm$^{-3}$, $\bar
E(k_{f0})= -15.3\,$MeV. The isospin-asymmetry energy has the value $A(k_{f0})= 
38.9\,$MeV at saturation density. The negative values of $S(k_f)$ indicate the
spin-instability of nuclear matter in this approximation.}  
\end{figure}

Now we can turn to numerical results. In Fig.\,2 we show the spin-asymmetry 
energy $S(k_f)$ of nuclear matter as a function the nucleon density $\rho= 
2k_f^3/3\pi^2$. The full line corresponds to a calculation up to fourth order 
in small momenta. It includes besides the kinetic energy term eq.(3) the
contributions from static $1\pi$-exchange and iterated $1\pi$-exchange. For 
reasons of consistency we have dropped the small relativistic 
$1/M^2$-correction in eq.(4) since it is of fifth order in the small momenta  
$k_f$ and $m_\pi$. The cut-off scale $\Lambda = 611\,$MeV has been adjusted to 
the nuclear matter saturation point $\rho_0= 0.173\,$fm$^{-3}$ and 
$\bar E(k_{f0})= -15.3\,$MeV \cite{lutzcontra}. The value of the nuclear 
matter compressibility, $K = k^2_{f0}\bar E''(k_{f0})= 252\,$MeV, as predicted
in this framework is consistent with a recent extrapolation from giant 
monopole resonances of heavy nuclei in ref.\cite{dario}, which gave $K = 
(260\pm 10)\,$MeV. One can read off from Fig.\,2 a negative value of the 
spin-asymmetry energy at saturation density: $S(k_{f0})=S(2m_\pi)=-20.1\,$MeV. 
It reveals the spin-instability of nuclear matter in this restricted 
framework. The small positive values of $S(k_f)$ at very low densities 
$\rho \leq 0.03\,$fm$^{-3}$ result from the incomplete cancellation between
the kinetic energy term eq.(3) proportional to $k_f^2$ and interaction 
contributions which grow in magnitude like $k_f^3$ (or higher powers of 
$k_f$). The largest negative contribution to $S(2m_\pi) = -20.1\,$MeV comes 
from the term linear in the cut-off $\Lambda$, eq.(13), which amounts to 
$-35.5\,$MeV at saturation density $k_{f0} = 2m_\pi$. Despite this 
overwhelming negative contribution to $S(k_f)$ one should not jump to the
conclusion that the use of cut-off regularization were the reason for the
spin-instability of nuclear matter in this calculation to fourth order. In
essence the cut-off $\Lambda$ is merely parameterizing the strength of a
zero-range (S-wave) NN-contact interaction with a fixed ratio of its
total-isospin $I=0$ and $I=1$ components. The isospin factor $(\vec \tau_1
\cdot \vec \tau_2)^2 = 3-2 \vec\tau_1 \cdot \vec \tau_2$ of the iterated
$1\pi$-exchange diagram determines this ratio as $9:1$. The quality of such a
fixed ratio can be tested via the isospin-asymmetry energy $A(k_f)$ (for
explicit expressions see section 3 in ref.\cite{nucmat}). The predicted value
$A(k_{f0})= 38.9\,$MeV for the isospin-asymmetry energy at saturation density
of the present fourth order calculation employing cut-off regularization
agrees within $10\%$ with a recent empirical determination, $A(k_{f0})=
(34\pm 2)\,$MeV \cite{dario}. The prescribed ratio $9:1$ between the $I=0$ and
$I=1$ components of the emerging contact-interaction works therefore rather 
well. A key observation in this context is that terms linear in the density 
$\rho$ satisfy the relation: $3S(k_f)_{\rm lin}+3A(k_f)_{\rm lin}+ 
2\bar E(k_f)_{\rm lin}=0$. From that one can conclude that even with two free 
parameters in the contact-interaction (adjusted to the empirical values of 
$\bar E(k_{f0})$ and $A(k_{f0})$) nuclear matter will be spin-unstable in a
calculation restricted to fourth order in small momenta. For example, in
dimensional regularization where the linear divergence $\int_0^\infty dl\,1$ is
set to zero one would instead employ a two-parameter momentum-independent 
NN-contact interaction. The adjustment to the empirical nuclear matter 
saturation point  $\bar E(k_{f0}=2m_\pi) =-15.3\,$MeV introduces then the
explicit term linear in density $\bar E(2m_\pi)_{\rm lin}  =-177.4\,$MeV while
the constraint from the empirical isospin-asymmetry energy $A(k_{f0})=
34\,$MeV requires in addition $A(2m_\pi)_{\rm lin} =148.9\,$MeV. These fixed
contributions imply via the abovementioned relation a term linear in density 
for the spin-asymmetry energy of $S(2m_\pi)_{\rm lin}  =-30.6\,$MeV. It is
almost as negative as the value $-35.5\,$MeV obtained in cutoff regularization
with only one adjusted cut-off scale $\Lambda$ and therefore the
spin-instability  of nuclear matter would still remain.

\begin{figure}
\begin{center}
\includegraphics[scale=1.,clip]{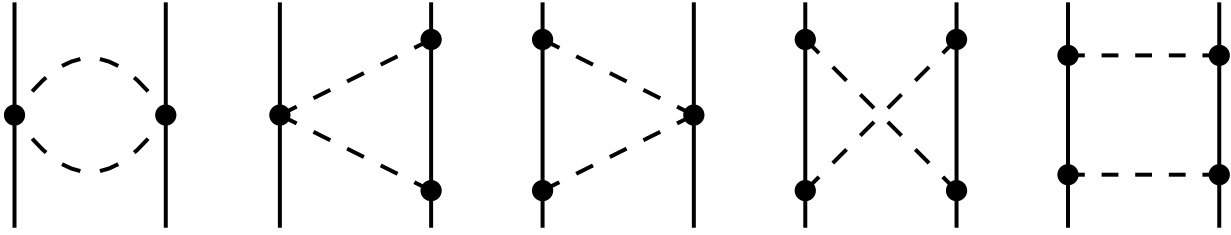}
\end{center}\vspace{-0.4cm}
\caption{One-loop diagrams of irreducible two-pion exchange between nucleons.}
\end{figure}

\begin{figure}
\begin{center}
\includegraphics[scale=0.994,clip]{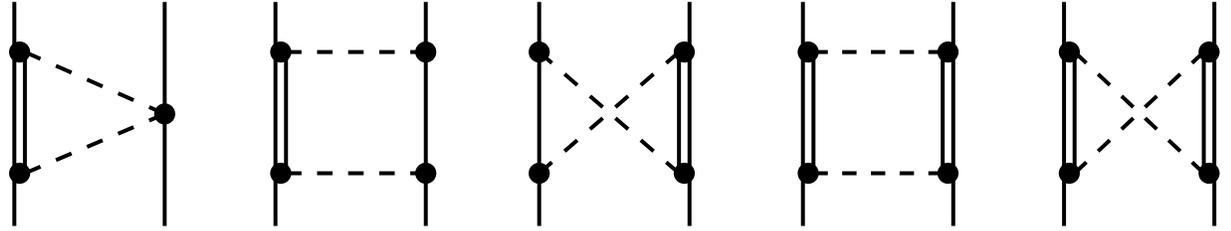}
\end{center}\vspace{-0.4cm}
\caption{One-loop two-pion exchange diagrams with single and double
$\Delta(1232)$-isobar excitation. Diagrams for which the role of both nucleons
is interchanged are not shown.}
\end{figure}

The foregoing discussion has shown that it is mandatory to include 
contributions of fifth order in the small momentum expansion in order to be
able to reach the spin-stability of nuclear matter. In the three-loop
approximation the terms of fifth order are generated by (irreducible)
two-pion-exchange between  nucleons. The corresponding one-loop diagrams for
elastic NN-scattering are shown in Fig.\,3. Since we are counting the 
delta-nucleon  mass splitting $\Delta= 293\,$MeV (together with $k_f$ and
$m_\pi$) as a small momentum scale the diagrams with single and double virtual
$\Delta(1232)$-isobar excitation shown in Fig.\,4 belong to the same order. 
The non-relativistic nucleon and delta propagators are both counted in this
scheme as an inverse small momentum. 

By closing the two open nucleon lines of the one-loop diagrams in Figs.\,3,4 
to either two or one ring one gets (in diagrammatic representation) the
Hartree or Fock contribution to the energy density of nuclear matter. The
Hartree contribution to the spin-asymmetry energy $S(k_f)$ vanishes
identically because the relevant $2\pi$-exchange NN T-matrix in forward 
direction is spin-independent \cite{gerst,nnpap}. The Fock contribution on the
other hand is obtained by integrating the spin- and isospin-contracted
T-matrix over the product of two Fermi spheres of radii $k_{\uparrow,
\downarrow} = k_f(1\pm \eta)^{1/3}$. We separate regularization dependent
short-range contributions to the T-matrix (originating from the ultra-violet
divergences of the one-loop diagrams in Figs.\,3,4) from the unique long-range
terms with the help of a twice-subtracted dispersion relation. The occurring
subtraction constants give rise to a contribution to the spin-asymmetry energy
of the form:      
\begin{equation} S(k_f)= (B_3-6B_{n,3}){k_f^3 \over 3 M^2} +  S_5 {k_f^5 \over
M^4}\,. \end{equation} 
The two dimensionless parameters $B_3 = -7.99$ and $B_{n,3}=-0.95$ have been
adjusted in ref.\cite{deltamat} to few empirical nuclear matter properties
(such as the maximal binding energy per particle $-\bar E(k_{f0})=16\,$MeV and 
the isospin-asymmetry energy $A(k_{f0})= 34\,$MeV with $k_{f0}=261.6\,$MeV). 
Again, we recognize in the first part of eq.(14) the relation $3S(k_f)_{\rm
lin}+3A(k_f)_{\rm lin}+2\bar E(k_f)_{\rm lin}=0$ for terms linear in density
$\rho= 2k_f^3/3\pi^2$. The other subtraction constant $S_5$ in front of the
$k_f^5/M^4$-term is (a priori) not constrained by any property of
spin-saturated nuclear matter.  The long-range parts of the $2\pi$-exchange
(two-body) Fock diagrams can be expressed as a dispersion-integral:  
\begin{eqnarray} S(k_f)&=&  {1 \over 36\pi^3} \int_{2m_\pi}^{\infty} \!\!
d\mu  \bigg\{{\rm Im}(V_C+3W_C)\bigg[3\mu k_f-{6k_f^3 \over \mu} +{16k_f^5
\over \mu^3} - {3\mu^3 \over 4k_f}\ln\bigg( 1+ {4k_f^2 \over \mu^2} \bigg)
\bigg] \nonumber \\ && + {\rm Im}(V_T+3W_T) \bigg[ 6\mu^3 k_f+ 4\mu k_f^3 
-{\mu^3 \over 2k_f} ( 8k_f^2+3\mu^2) \ln\bigg( 1+{4k_f^2 \over \mu^2}\bigg)
\bigg]\bigg\}\,, \end{eqnarray}
where Im$V_C$, Im$W_C$, Im$V_T$ and Im$W_T$ are the spectral functions of the
isoscalar and isovector central and tensor NN-amplitudes, respectively. 
Explicit expressions of these imaginary parts for the contributions of the 
triangle diagram with single $\Delta$-excitation and the box diagrams with 
single and double $\Delta$-excitation can be easily constructed from the 
analytical formulas given in section 3 of ref.\cite{gerst}. The $\mu$- and 
$k_f$-dependent weighting functions in eq.(15) take care that at low and
moderate densities this spectral-integral is dominated by low invariant
$\pi\pi$-masses $2m_\pi< \mu <1\,$GeV. The contributions to the spin-asymmetry
energy $S(k_f)$ from irreducible $2\pi$-exchange (with only nucleon 
intermediate states, see Fig.\,3) can also be cast into the form eq.(15). The 
corresponding non-vanishing spectral functions read \cite{nnpap}: 
\begin{equation} {\rm Im}W_C(i\mu) = {\sqrt{\mu^2-4m_\pi^2} \over 3\pi 
\mu (4f_\pi)^4} \bigg[ 4m_\pi^2(1+4g_A^2-5g_A^4) +\mu^2(23g_A^4-10g_A^2-1) + 
{48 g_A^4 m_\pi^4 \over \mu^2-4m_\pi^2} \bigg] \,, \end{equation}
\begin{equation} {\rm Im}V_T(i\mu) = - {6 g_A^4 \sqrt{\mu^2-4m_\pi^2} \over 
\pi  \mu (4f_\pi)^4}\,. \end{equation}

\begin{figure}
\begin{center}
\includegraphics[scale=1.0,clip]{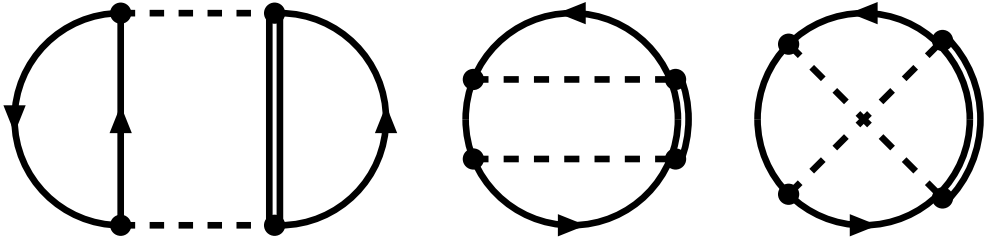}
\end{center}\vspace{-0.4cm}
\caption{Hartree and Fock three-body diagrams related to $2\pi$-exchange with 
single virtual $\Delta$-isobar excitation. They represent interactions between
three nucleons in the Fermi sea. In the case of isospin-symmetric nuclear 
matter the isospin factors of these diagrams are 8, 0, and 8, in the 
order shown. The combinatoric factor is 1 for each diagram.} 
\end{figure}

Next, we come to the additional $2\pi$-exchange three-body terms which arise 
from Pauli blocking of intermediate nucleon states (i.e. from the $-2\pi 
\theta(k_{\uparrow,\downarrow} -|\vec p\,|)$ terms in the in-medium nucleon 
propagators \cite{nucmat}). The corresponding closed Hartree and Fock diagrams 
with single virtual $\Delta$-excitation are shown in Fig.\,5. For 
isospin-symmetric nuclear matter their isospin factors are 8, 0, and 8, in the 
order shown. The contribution of the left three-body Hartree diagram to the 
spin-asymmetry energy $S(k_f)$ has the following analytical form:  
\begin{equation} S(k_f)={g_A^4 m_\pi^6 u^2\over 9\Delta(2\pi f_\pi)^4} \bigg[
\bigg({9\over 4}+4u^2\bigg)\ln(1+4u^2)-2u^4(1+3\zeta) -8u^2-{u^2\over 1+4u^2}
\bigg] \,. \end{equation}
The delta propagator shows up in this expression merely via the (reciprocal)
mass-splitting $\Delta =293\,$MeV. Furthermore, we have already inserted in 
eq.(18) the empirically well-satisfied relation $g_{\pi N\Delta} = 3g_{\pi
N}/\sqrt{2}$ for the $\pi N\Delta$-coupling constant. The parameter $\zeta = 
-3/4$ has been introduced in section 2 of ref.\cite{deltamat} in order 
to reduce a too strongly repulsive $\rho^2$-term in the energy particle $\bar
E(k_f)$. It controls the strength of a three-nucleon contact interaction
$\sim (\zeta g_A^4/\Delta f_\pi^4) \, (\bar NN)^3$ which has the interesting 
property that it contributes equally but with opposite sign to the energy per 
particle $\bar E(k_f)$ and the spin-asymmetry energy $S(k_f)$. The
contribution of the right three-body Fock diagram in Fig.\,5 to the
spin-asymmetry energy $S(k_f)$ can be represented as:   
\begin{eqnarray} S(k_f)&=&{g_A^4 m_\pi^6 \over 36\Delta(4\pi f_\pi)^4u^3}
\int_0^u\!\! dx\Big\{ 4G_{S01}G_{S10}-2G_{S01}^2-6G_{S10}^2 \nonumber \\ && +
2 G_S (3G_S+8 G_{S01}-3 G_{S02}+2 G_{S11}-3 G_{S20})+ 2G_{T01}G_{T10}
\nonumber \\ && -7G_{T01}^2-3G_{T10}^2  + G_T(3G_T+8 G_{T01}-3G_{T02}
+2G_{T11} -3 G_{T20}) \Big\} \,,\end{eqnarray}
with the two auxiliary functions:
\begin{eqnarray} G_S&=& {4ux \over 3}( 2u^2-3) +4x\Big[
\arctan(u+x)+\arctan(u-x)\Big] \nonumber \\ && + (x^2-u^2-1) \ln{1+(u+x)^2
\over  1+(u-x)^2} \,,\end{eqnarray}
\begin{eqnarray} G_T &=& {ux\over 6}(8u^2+3x^2)-{u\over
2x} (1+u^2)^2  \nonumber \\ && + {1\over 8} \bigg[ {(1+u^2)^3 \over x^2} -x^4 
+(1-3u^2)(1+u^2-x^2)\bigg] \ln{1+(u+x)^2\over  1+(u-x)^2} \,.\end{eqnarray}
The double-indices on $G_S$ and $G_T$ have the same meaning as explained in 
eq.(11) for the function $G$. It is remarkable that the three-body Hartree and
Fock diagrams with virtual $\Delta$-excitation (shown in Fig.\,5) lead both to 
identical contributions to the spin-asymmetry energy $S(k_f)$ and the 
isospin-asymmetry energy $A(k_f)$ (compare eqs.(18,19) here with eqs.(48,49) 
in ref.\cite{deltamat}). This feature originates from the specific  
spin-isospin structure of the (non-relativistic) $\pi N \Delta$-transition
vertex \cite{gerst}.

\begin{figure}
\begin{center}
\includegraphics[scale=0.7]{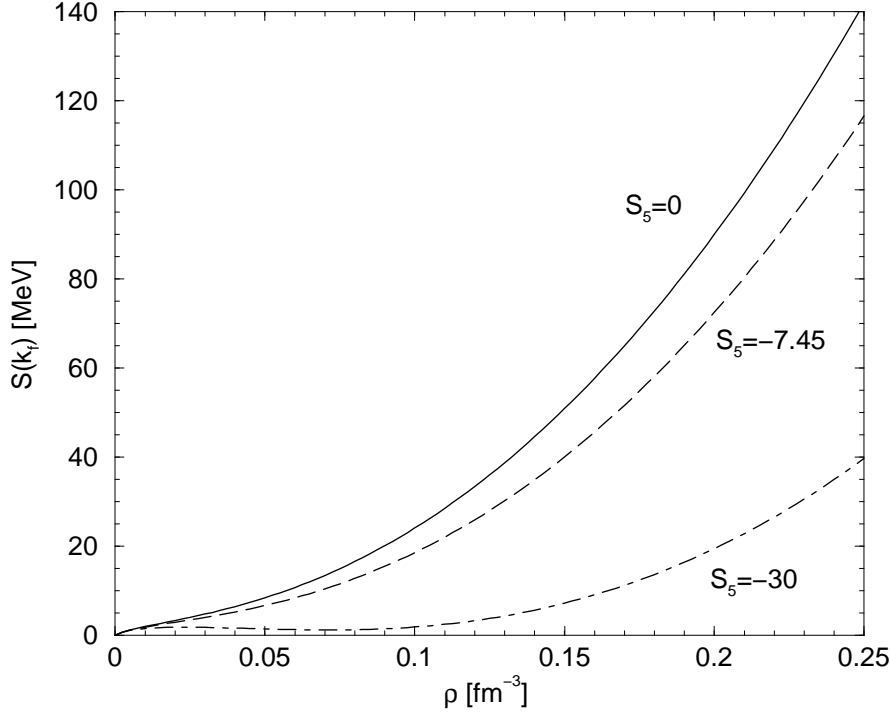}
\end{center}\vspace{-.8cm}
\caption{The spin-asymmetry energy $S(k_f)$ of nuclear matter versus the 
nucleon density $\rho= 2k_f^3/3\pi^2$. In comparison to Fig.\,2 the effects
from $2\pi$-exchange with single and double virtual $\Delta$-isobar excitation
are now included. The full, dashed and dashed-dotted curves correspond to the
choices $S_5=0,\,-7.45$ and $-30$ of the short-range parameter $S_5$. The 
positive values of $S(k_f)$ ensure the spin-stability of nuclear matter.}
\end{figure}

In Fig.\,6 we show again the spin-asymmetry energy $S(k_f)$ of nuclear matter 
as a function of the nucleon density $\rho= 2k_f^3/3\pi^2$. The full line 
includes all the contributions from chiral $1\pi$- and $2\pi$-exchange written
down in eqs.(3-9,14-19). The (yet undetermined) short-range parameter $S_5$ 
has been set to zero, $S_5=0$. We note as an aside that the term linear in the
density and the cut-off $\Lambda$ eq.(13) is of course now not counted extra 
since the parameter $B_3-6 B_{n,3}=-2.29$ \cite{deltamat} collects all such 
possible terms. One observes in Fig.\,6 a positive spin-asymmetry energy 
$S(k_f)$ which rises monotonically with the density $\rho$. The inclusion of 
the chiral $\pi N\Delta$-dynamics has made nuclear matter spin-stable. This
result is of utmost importance since it assures that the improved nuclear
matter properties reported in ref.\cite{deltamat} are based on a spin-stable
ground state. It is also interesting to look at numerical values of $S(k_f)$ 
and their decomposition. At a Fermi momentum of $k_f = 2m_\pi$ (corresponding 
to $\rho = 0.173\,$fm$^{-3}$) the spin-asymmetry energy is now $S(2m_\pi)= 
67.3\,$MeV (setting $S_5=0$). The most significant changes in comparison to 
the previous fourth order calculation come from the term linear in density 
which gets reduced (by a factor of $2.1$) to $-17.0\,$MeV and the two-body 
Fock and three-body Hartree contributions eqs.(15,18) which amount together  
to $54.6\,$MeV + $13.7\,$MeV = $68.3\,$MeV. We note also that about one 
quarter thereof (namely $16.2\,$MeV) stem from the three-body contact 
interaction proportional to $\zeta=-3/4$.   

The remaining open question concerns the size of the short-distance parameter
$S_5$ in eq.(14). Large negative values could again endanger the 
spin-stability of nuclear matter. In order to get an estimate of $S_5$ we
bring into play the complete set of four-nucleon contact-couplings written
down in eqs.(3,4) of ref.\cite{evgeni}. This set represents the most general
short-range NN-interaction quadratic in momenta and it involves seven
low-energy constants $C_1, \dots, C_7$. After computing the spin-asymmetry
energy $S(k_f)$ from the corresponding contact-potential in Hartree-Fock
approximation we find:      
\begin{eqnarray} S_5 &=& {M^4 \over 18\pi^2} (C_2-4C_1+3C_4+C_7) \nonumber \\
&=&  {M^4 \over 96\pi^3} \Big[ 2C(^3\!S_1)-2C(^1\!S_0)- C(^1\!P_1)+C(^3\!P_0)
+3C(^3\!P_1)+5 C(^3\!P_2)\Big] \,. \end{eqnarray}   
In the second line of eq.(22) we have reexpressed the relevant linear 
combination of $C_{1,2,4,7}$ through the so-called spectroscopic low-energy
constants which characterize the short-range part of the NN-potential in the
spin-singlet and spin-triplet $S$- and $P$-wave states. In that representation
we obtain from the entries of table IV in ref.\cite{evgeni} for the three 
high-precision NN-potentials CD-Bonn \cite{cdbonn}, Nijm-II \cite{nijm} and
AV-18 \cite{av18} the numbers $S_5 = -7.10,\, -7.98$ and $-7.27$. The dashed
line in Fig.\,6 shows the spin-asymmetry energy $S(k_f)$ which results from
taking their average value $S_5= -7.45$. The effective reduction of the 
spin-asymmetry energy by about $20\%$ is harmless and the spin-stability of 
nuclear matter is equally well guaranteed. The dashed-dotted curve in Fig.\,6 
corresponds to the extreme choice $S_5 = -30$. One can see that even with such 
a large negative $S_5$-value, which exceeds the estimate from realistic 
NN-potentials by a factor of 4, the spin-stability of nuclear matter still 
remains preserved. We can therefore conclude that with the inclusion of the 
chiral $\pi N\Delta$-dynamics spin-stability becomes a robust property of 
nuclear matter. This is an important finding and it goes conform with the 
improved isospin properties observed in ref.\cite{deltamat}.

In summary we have investigated in this work the spin-stability of nuclear
matter in the framework of chiral perturbation theory. For that purpose we 
have calculated the density-dependent spin-asymmetry energy $S(k_f)$ of 
isospin-symmetric nuclear matter to three-loop order. The interaction 
contributions to $S(k_f)$ originate from one-pion exchange, iterated one-pion 
exchange, and (irreducible) two-pion exchange with no, single, and double 
virtual $\Delta$-isobar excitation. We have found that the truncation to 
$1\pi$-exchange and iterated $1\pi$-exchange terms (which leads already to a 
good nuclear matter equation of state) is spin-unstable, since $S(k_{f0})<0$. 
This statement holds independently of the regularization scheme if the contact
terms (generating contributions linear in density) are consistent with the  
empirical nuclear matter bulk properties: $\bar E(k_{f0}) \simeq -16\,$MeV and
$A(k_{f0}) \simeq 34\,$MeV. The inclusion of the chiral $\pi N\Delta$-dynamics
on the other hand guarantees the spin-stability of nuclear matter. The 
corresponding spin-asymmetry energy $S(k_f)$ stays positive within a wide
range of an undetermined short-range parameter $S_5$, which we have also
estimated from realistic NN-potentials. Our results reemphasize the important
role played by two-pion exchange with  virtual $\Delta$-isobar excitation for
the nuclear matter many-body problem.


\begin{thebibliography}{99}
\bibitem{lutz} M. Lutz, B. Friman, and Ch. Appel, {\it Phys. Lett.} {\bf
B474}, 7 (2000).\vs
\bibitem{nucmat} N. Kaiser, S. Fritsch and W. Weise, \textit{Nucl. Phys.} 
\textbf{A697}, 255 (2002); and references therein.\vs
\bibitem{deltamat} S. Fritsch, N. Kaiser, and W. Weise, ''Chiral approach to
nuclear matter: Role of two-pion exchange with virtual delta-isobar 
excitation'', nucl-th/0406038; submitted to \textit{Nucl. Phys.} \textbf{A}.\vs
\bibitem{gerst} N. Kaiser, S. Gerstend\"orfer and W. Weise, \textit{Nucl. 
Phys.} \textbf{A637}, 395 (1998).\vs
\bibitem{pot} N. Kaiser, S. Fritsch and W. Weise, \textit{Nucl. Phys.} 
\textbf{A700}, 343 (2002).\vs
\bibitem{liquidgas} S. Fritsch, N. Kaiser and W. Weise, \textit{Phys. Lett.} 
\textbf{B545}, 73 (2002).\vs
\bibitem{lutzcontra}  S. Fritsch and N. Kaiser, \textit{Eur. Phys. J.} 
\textbf{A17}, 11 (2003).\vs
\bibitem{chang} B.D. Chang, \textit{Phys. Lett.} \textbf{B56}, 205 (1975).\vs
\bibitem{baeckman} S.O. B\"ackman, A.D. Jackson and J. Speth, \textit{Phys. 
Lett.} \textbf{B56}, 209 (1975).\vs
\bibitem{krewald} S. Krewald, V. Klemt, J. Speth and A. Faessler, 
\textit{Nucl. Phys.} \textbf{A281}, 166 (1977).\vs
\bibitem{braghin} F.L. Braghin, {\it Int. J. Mod. Phys.} {\bf E12}, 755
(2003); and references therein.\vs 
\bibitem{pavan} M.M. Pavan et al., {\it Phys. Scr.} {\bf T87}, 65 (2000).\vs
\bibitem{dario} D. Vretenar, T. Niksic and P. Ring, {\it Phys. Rev.} {\bf
C68}, 024310 (2003).\vs
\bibitem{nnpap} N. Kaiser, R. Brockmann and W. Weise, \textit{Nucl. Phys.} 
\textbf{A625}, 758 (1997).\vs
\bibitem{evgeni} E. Epelbaum, Ulf-G. Mei{\ss}ner, W. Gl\"ockle, and C. Elster, 
{\it Phys. Rev.} {\bf C65}, 044001 (2002); and references therein.\vs
\bibitem{cdbonn} R. Machleidt,  {\it Phys. Rev.} {\bf C63}, 024001 (2001); and
references therein.\vs
\bibitem{nijm} V.G.J. Stoks, R.A.M. Klomp, C.P.F. Terheggen and J.J. de Swart,
{\it Phys. Rev.} {\bf C49}, 2950 (1994).\vs
\bibitem{av18} R.B. Wiringa, V.G.J. Stoks and R. Schiavilla, {\it Phys. Rev.} {\bf C51}, 38 (1995).\vs
\end{thebibliography}
\end{document}